\documentclass[twocolumn,showpacs,amsmath,amssymb]{revtex4}
\usepackage{graphicx}% Include figure files
\usepackage{dcolumn}% Align table columns on decimal point
\usepackage{bm}% bold math
\newcommand{\etal}{\emph{et al.}}
\newcommand{\be}{\begin{equation}}
\newcommand{\ee}{\end{equation}}
\newcommand{\bfig}{\begin{figure}}
\newcommand{\efig}{\end{figure}}
\newcommand{\incl}{\includegraphics}
\begin{document}      

\title{Anomalous-Hall heat current and Nernst effect in the ferromagnet CuCr$_2$Se$_{4-x}$Br$_x$.\footnote{Phys. Rev. Lett. {\bf %%@
93}, 226601 (2004), }} 
\author{Wei-Li Lee$^{1}$, S. Watauchi$^{2,*}$, V. L. Miller$^2$, R. J. Cava$^{2}$, and N. P. Ong$^{1}$}      
\affiliation{
Departments of Physics$^1$ and Chemistry$^2$, Princeton University, New Jersey 08544, U.S.A. 
}

\date{\today}      

\begin{abstract}
In a ferromagnet, an anomalous-Hall heat current, given by the off-diagonal Peltier term $\alpha_{xy}$, accompanies the anomalous Hall current.  By %%@
combining Nernst, thermopower and Hall experiments, we have measured how $\alpha_{xy}$ varies with hole density and lifetime $\tau$ in $\rm %%@
CuCr_2Se_{4-x}Br_x$.  At low temperatures $T$, we find that $\alpha_{xy}$ is independent of $\tau$, consistent with anomalous-velocity theories.  Its %%@
magnitude is fixed by a microscopic geometric area ${\cal A}\sim 34\; \mathrm{\AA}^2$.  Our results are incompatible with some models of the Nernst %%@
effect in ferromagnets.
\end{abstract}
\pacs{72.15.Gd,72.15.Jf,75.47.-m,72.10.Bg}
\maketitle                   % Produces the title
In a ferromagnet, the anomalous Hall effect (AHE) is the appearance of a spontaneous Hall current flowing parallel to $\bf E\times M$ where $\bf E$ is the %%@
electric field and $\bf M$ the magnetization~\cite{Hurd}.  Karplus and Luttinger (KL)~\cite{Karplus} proposed that the AHE current originates from an %%@
anomalous velocity term which is non-vanishing in a ferromagnet.  The topological nature of the KL theory has been of considerable interest %%@
recently~\cite{Niu,Nagaosa,Jungwirth,Murakami}.  Experimentally, strong evidence for the dissipationless nature of the AHE current has been obtained in %%@
the spinel ferromagnet $\rm CuCr_2Se_{4-x}Br_x$.   Lee \etal~\cite{Lee} reported that, despite a 1000-fold increase in the resistivity $\rho$ induced by %%@
varying the Br content $x$, the anomalous Hall conductivity (normalized per carrier and measured at 5 K) stays at the same value, in agreement with KL's %%@
prediction.  A test of the anomalous velocity theory against the AHE in Fe has also been reported~\cite{Yao}.

It has long been known that an anomalous heat current density ${\bf J}^Q$ also accompanies the AHE current in the absence of any temperature %%@
gradient~\cite{Smith,Handbook}.  In principle, ${\bf J}^Q$ can provide further information on the origin of the AHE, but almost nothing is known about its %%@
properties.  A weak heat current is a challenge to measure.  Instead, one often performs the `reciprocal' Nernst experiment in which a temperature %%@
gradient $-\mbox{\boldmath$\nabla$}T$ produces a transverse charge current, which is detected as a Nernst electric field ${\bf E}_N$ parallel to ${\bf %%@
M}\times (-\mbox{\boldmath$\nabla$} T)$.  However, in previous Nernst experiments on ferromagnets~\cite{Smith,Handbook,Kondorskii}, $J^Q$ was not %%@
found because other transport quantities were not measured.  Combining the Nernst signal with the AHE resistivity $\rho'_{xy}$ and the thermopower, we %%@
have determined how the transport quantity $\alpha_{xy}$ relevant to $J^Q$ varies in $\rm CuCr_2Se_{4-x}Br_x$ as the hole density $n_h$ and carrier %%@
lifetime $\tau$ are greatly changed under doping.  We show that $\alpha_{xy}$ has a strikingly simple form, with its magnitude scaled by a microscopic %%@
geometric area $\cal A$.

We apply a gradient $-\mbox{\boldmath$\nabla$} T||\hat{\bf x}$ to an electrically-isolated sample in a magnetic field $\bf H||\hat{z}$.  Along $\hat{\bf x}$, the %%@
charge current driven by $-\mbox{\boldmath$\nabla$} T$ is balanced by a backflow current produced by a large $E_x$ which is detected as the %%@
thermopower $S = E_x/|\nabla T|$.  Along the transverse direction $\hat{\bf y}$, however, both $E_x$ and $-\mbox{\boldmath$\nabla$} T$ generate %%@
Hall-type currents.  In general, the charge current in the presence of $\bf E$ and $-\mbox{\boldmath$\nabla$} T$ is
${\bf J} = \mbox{\boldmath $\sigma$}{\bf\cdot E} + \mbox{\boldmath$\alpha$}\cdot (-\mbox{\boldmath$\nabla$} T)$
with \mbox{\boldmath$\sigma$} and \mbox{\boldmath$\alpha$} the electrical and thermoelectric (`Peltier') conductivity tensors, respectively.  Setting $J_y %%@
= 0$, we obtain the Nernst signal $e_N\equiv E_y/|\nabla T| =  \rho \alpha_{xy} + \rho_{xy}\alpha$ where $\alpha\equiv\alpha_{xx}$~\cite{Ong}.  Hence, as %%@
noted, the Nernst signal results from the 2 distinct $y$-axis currents $\alpha_{yx}(-\mbox{\boldmath$\nabla$} T)$ and $\sigma_{yx}E_x$.  In a ferromagnet, %%@
the former is our desired gradient-driven current, whereas the latter comprises the `dissipationless' AHE current and the weak %%@
ordinary Hall current.  

In terms of the thermopower $S = \rho\alpha$ and Hall angle $\tan\theta_H = \rho_{yx}/\rho$, we may express $\alpha_{xy}$ as
\be
\rho\alpha_{xy} = e_N + S\tan\theta_H.
\label{axy}
\ee
Hence, to find $\alpha_{xy}$, we need to measure $e_N$, $S$, $\rho_{xy}$ and $\rho$.  Knowing  $\alpha_{xy}$, we readily find the transverse heat %%@
current $J^Q_y = \tilde\alpha_{yx}E_x$ since $\tilde\alpha_{yx} = \alpha_{yx}T$ by Onsager reciprocity.   

The spinel $\rm CuCr_2Se_4$ is a conducting ferromagnet with a Curie temperature $T_C\sim$ 450 K.  Because the exchange between local moments %%@
in Cr is mediated by superexchange through 90$^\mathrm{o}$ Cr-Se-Cr bonds rather than the carriers, $T_C$ is not significantly reduced even when the %%@
hole population $n_h$ drops by a factor of 30 under Br doping ($M$ at 5 K actually increases by 20$\%$)~\cite{Miyatani,Lee}.  Using iodine vapor %%@
transport, we have grown crystals with $x$ from 0.0 to 1.0.  As $x$ increases from 0 to 1, the value of $\rho$ at 5 K increases by $\sim 10^3$, while %%@
$\rho'_{xy}/n_h$ increases by $\sim 10^6$~\cite{Lee}.  The tunability of $n_h$ and the robustness of $M$ under doping make this system attractive for %%@
studying charge transport in a lattice with broken time-reversal symmetry.  The behavior of $\rho$, $M$, and $\rho'_{xy}$ vs. $x$ are described in %%@
Ref.~\cite{Lee}.  
\bfig[h]			% Fig 1
\incl[width=9cm]{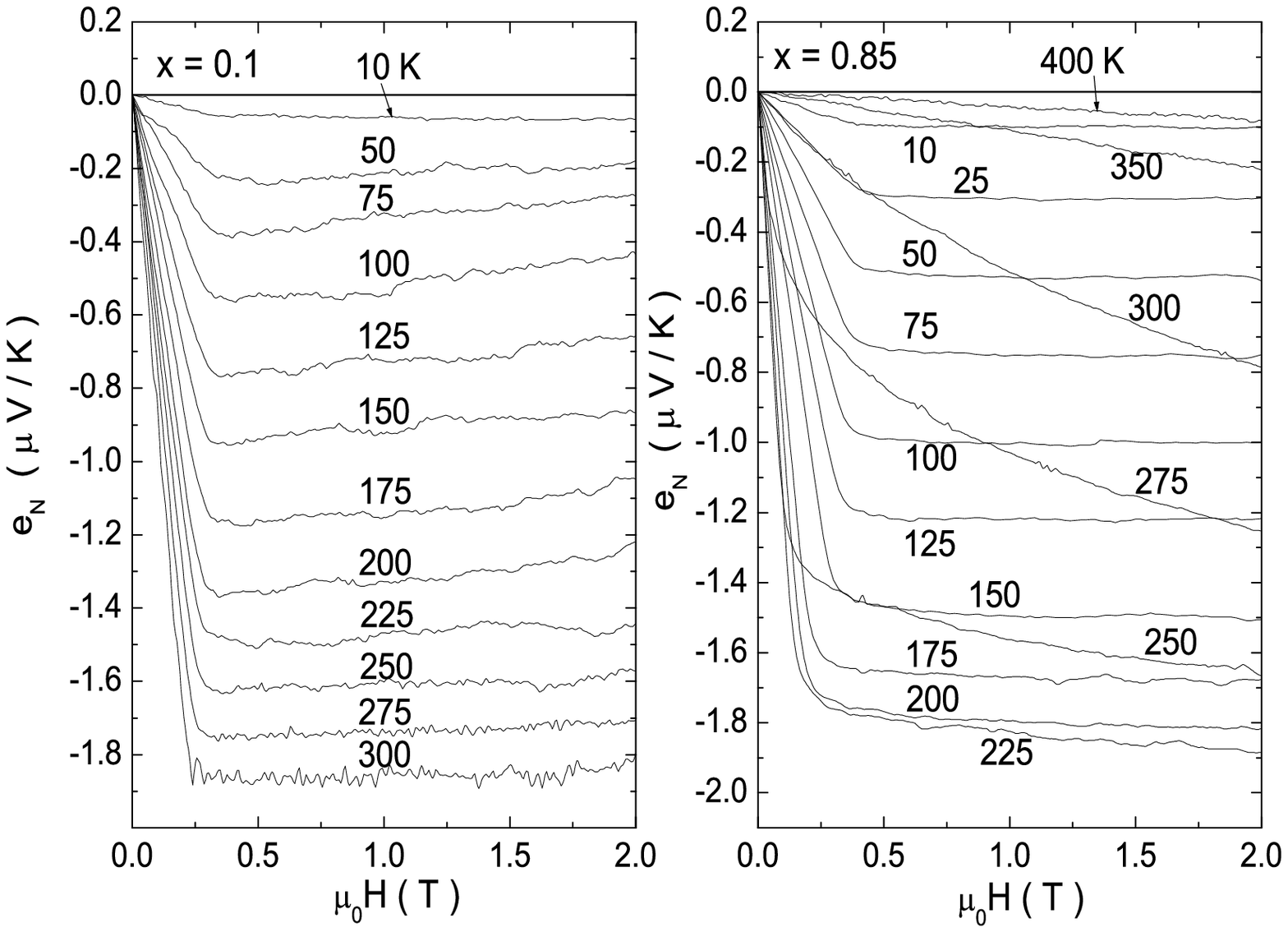}
\caption{\label{NH}  Curves of the measured $e_N = E_y/|\nabla T|$ vs. $H$ in $\rm CuCr_2Se_{4-x}Br_x$ with $x$ = 0.1 (left panel) and 0.85 (right).  In %%@
the ferromagnetic state below $T_C$, $e_N$ saturates to a constant when $H$ exceeds $H_s$ reflecting the $M$-$H$ curve.  The scaling factor $Q_s$ %%@
increases rapidly as $T$ increases from 10 K to $T_C$.  In the right panel, $e_N$ continues to scale as the $M$-$H$ curve in the paramagnetic regime %%@
(275-400 K). 
}
\efig

Figure \ref{NH} shows profiles of $e_N$ vs. $H$ at selected $T$ in 2 samples with $x$ = 0.1 and 0.85 and $T_C$ = 400 and 275 K, respectively.  As %%@
noted above, $e_N(T,H)$ is the sum of two terms both of which scale as $M$.  The magnitude $|e_N|$ initially increases as $H$ rotates domains into %%@
alignment and then saturates to a constant for $H>H_s$, the saturation field.  The sign of $e_N$ -- negative in all samples -- reflects the sign of the %%@
dominant term~\cite{sign}.  

In the sample with $x$ = 0.85, the curves above $T_C$ show that the scaling also holds in the paramagnetic regime where the susceptibility has the %%@
Curie-Weiss form $\chi\sim 1/(T-T_C)$ in weak $H$.  In analogy with the Hall resistivity $\rho_{xy} = R_0\mu_0H + R_s\mu_0M$, with $R_0$ and $R_s$ %%@
the ordinary and anomalous Hall coefficients, respectively, it is customary to express the scaling between the $e_N$-$H$ and $M$-$H$ %%@
curves by writing
\be
e_N = Q_0\mu_0 H + Q_s\mu_0 M.
\label{eN}
\ee
For $T<T_C$ in all samples, the $Q_0$ term cannot be resolved, so that $e_N\simeq Q_s\mu_0 M$.  Moreover, below 50 K, $M$ changes only weakly %%@
with $x$ (by 20$\%$ over the whole doping range), so that the saturated value of the Nernst signal $e_N^{sat}$ differs from $Q_s$ by a factor that is only %%@
weakly $x$ dependent.  

\bfig[h]			% Fig 2
\incl[width=6cm]{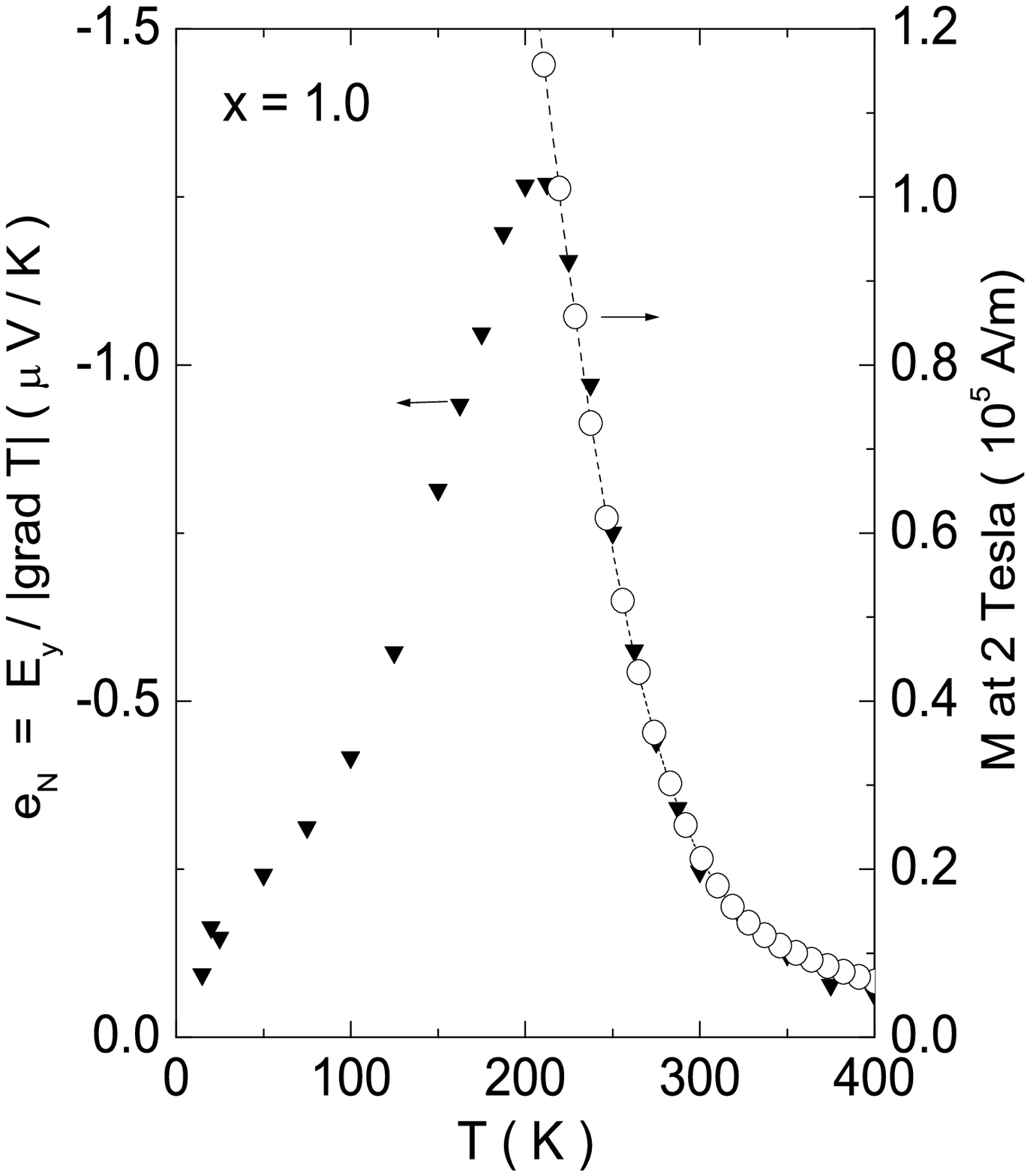}
\caption{\label{eN}  The $T$ dependence of the Nernst signal $e_N$ (solid triangles) measured at 2 T in the sample with $x$ = 1.0.  Above $T_C$, $e_N$ %%@
is compared with the paramagnetic magnetization $M$ at 2 T (open circles).   
}
\efig

The Nernst signal has very different characteristic behaviors below and above $T_C$.  As an example, Fig. \ref{eN} shows $e_N^{sat}$ measured at 2 T %%@
in the sample with $x$ = 1.0 ($T_C$ = 210 K).  Between 5 and 100 K, $e_N^{sat}$ increases linearly with $T$.  Above 100 K, $e_N^{sat}$ rises more %%@
steeply to a sharp peak 200 K, and then falls steeply above $T_C$.  As noted, in the paramagnetic regime, the Nernst signal matches the behavior of $M$ %%@
as a function of both $T$ and $H$.  Figure \ref{eN} shows that the $T$ dependence of $e_N$ closely follows that of $M = \chi H$ (both are measured at 2 %%@
T).  The experiment shows that, in a gradient, fluctuations of the paramagnetic magnetization lead to a significant transverse electrical current that is %%@
proportional to the average magnetization (this has not been noted before, to our knowledge).  We express the proportionality as 
\be
\alpha_{xy} = \beta M \quad(T>T_C),
\label{beta}
\ee
where $\beta$ is only weakly $T$ dependent (it decreases by $5\%$ between 250 and 400 K).  The parameter $\beta$ plays the important role of relating %%@
the magnitudes of the paramagnetic $M$ and the transverse electronic current (through the Nernst signal).   Its very small value ($\beta\simeq 2\times %%@
10^{-7}$ K$^{-1}$ at 250 K) reflects the strikingly weak coupling between the fluctuating $M$ and $e_N$ in ferromagnets, compared with vortex flow in a %%@
superconductor~\cite{Ong,vortex}.   With the growth of long-range magnetic order below $T_C$, Eq. \ref{beta} ceases to be valid.

In the ferromagnetic state, we restrict our attention to the regime below 100 K where $e_N^{sat}$ is nominally linear in $T$.  Figure \ref{Stan}a shows %%@
curves in this regime for the 5 samples studied.  The slopes of the low-$T$ curves are not monotonic in $x$.  As $x$ is increased from 0.1, the slope %%@
attains a maximum value at $x$ = 0.25, and then decreases to a value close to its initial value when $x$ reaches 1.0.  This is perhaps unsurprising since %%@
$e_N$ involves transport quantities $S$, $\rho$ and $\rho_{xy}$ with opposite trends vs. $x$.  By Eq. \ref{axy}, we may find the curve of $\alpha_{xy}$ vs. %%@
$T$ by adding the curves $e_N$ and $S\tan\theta_H$ (Fig. \ref{Stan}b) and dividing by $\rho$.  The data in Figs. \ref{Stan}a and \ref{Stan}b show that %%@
these terms are opposite in sign for $x>$ 0.3.  With increasing $x$, their mutual cancellation suppresses $\alpha_{xy}$ strongly.  In particular, at the largest %%@
$x$ (0.85 and 1.0), the cancellation is nearly complete and $\alpha_{xy}$ is very small below 100 K, i.e. the observed $e_N$ is nearly entirely from the %%@
AHE of the backflow current.  For $0.1\le x<0.3$, $S\tan\theta_H$ is negligible and $e_N$ largely reflects the behavior of $\alpha_{xy}$.  We exclude from %%@
our study the undoped compound $\rm CuCr_2Se_4$ because $e_N$ and $\rho'_{xy}$ were not resolved at low $T$.  These trends emphasize the %%@
importance of knowing all 4 transport quantities, instead of just $e_N$, to discuss ${\bf J}^Q$ meaningfully.
\bfig[h]			% Fig 3
\incl[width=9cm]{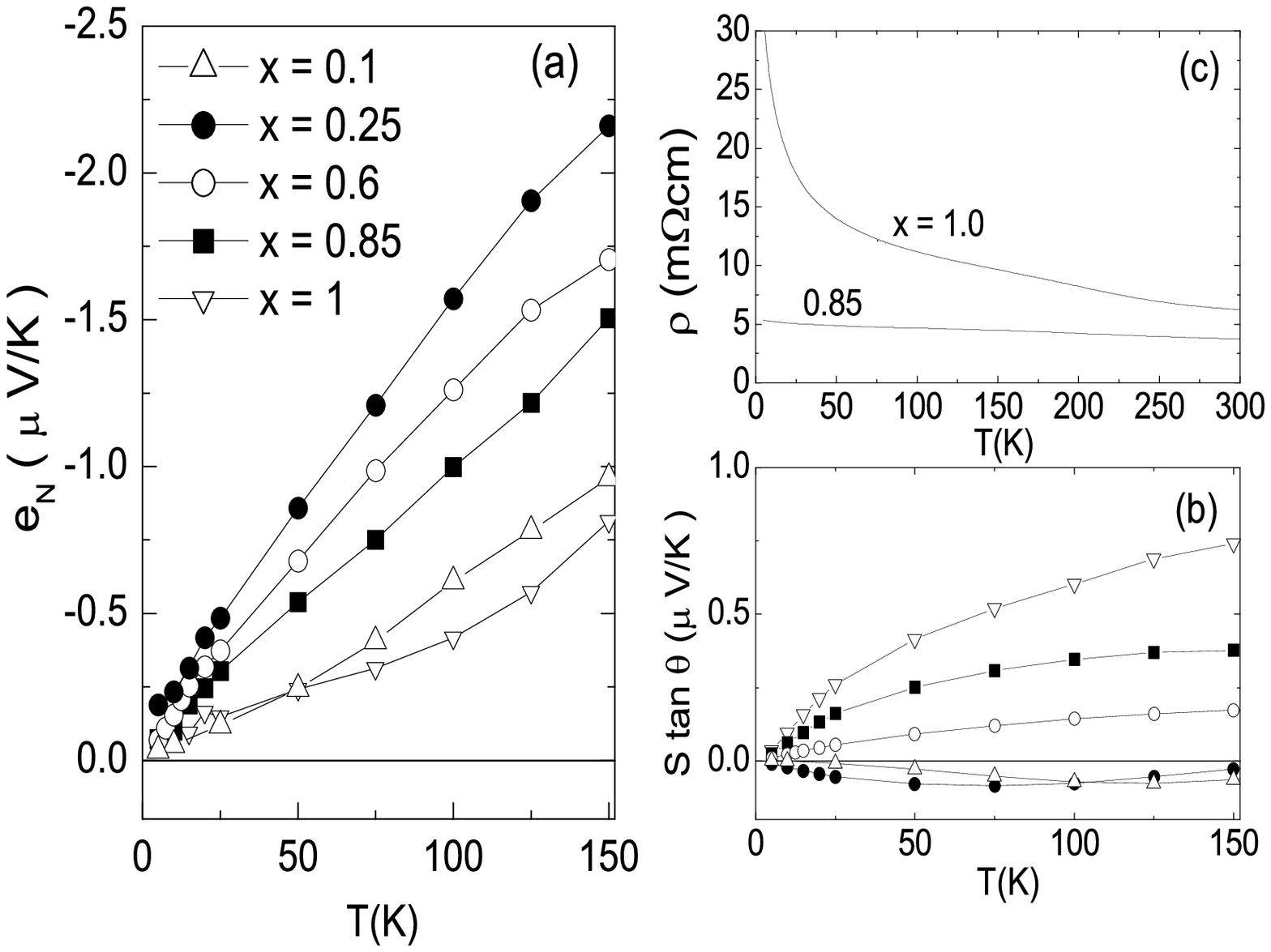}
\caption{\label{Stan}  (Panel a)  Curves of $e_N$ vs. $T$ below 150 K in 5 samples with doping $0.1\le x \le 1.0$ showing nominal $T$-linear behavior at %%@
low $T$.  The slopes vary non-monotonically with $x$.  Panel (b) shows the Hall-current term $S\tan\theta_H$ measured in the same samples.  For $x>$ %%@
0.3, $S\tan\theta_H$ is opposite in sign from $e_N$ [the symbol key applies to both (a) and (b)].  Panel (c) shows the sharp change in the $\rho$-$T$ %%@
profiles in the samples with $x$ = 0.85 and 1.0.  At low $T$, $\rho$ at 0.85 is metallic, but at 1.0, $\rho$ reveals hopping between %%@
strongly localized states.
}
\efig

\bfig[h]			% Fig 4
\incl[width=9cm]{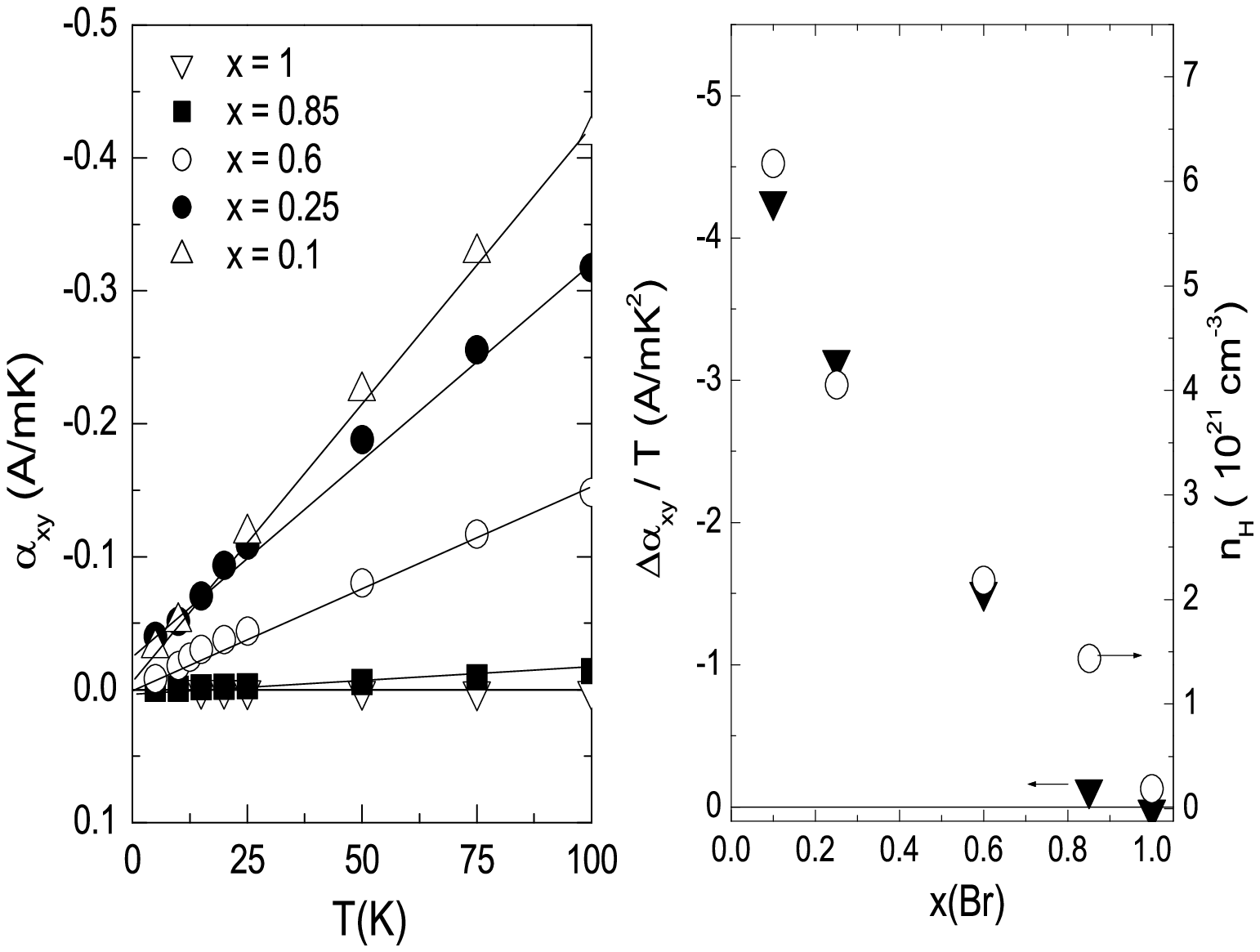}
\caption{\label{axyT} [Panel (a)] Curves of $\alpha_{xy}$ vs. $T$ obtained by adding $e_N$ and $S\tan\theta_H$ (Eq. \ref{axy}).  The slope $b(x)$ now %%@
falls monotonically as $x$ increases to 1.0.  Panel (b) compares how the slope $b(x) = \Delta\alpha_{xy}/T$ and $n_h$ vary with $x$.
}
\efig
\bfig[h]			% Fig 5
\incl[width=6cm]{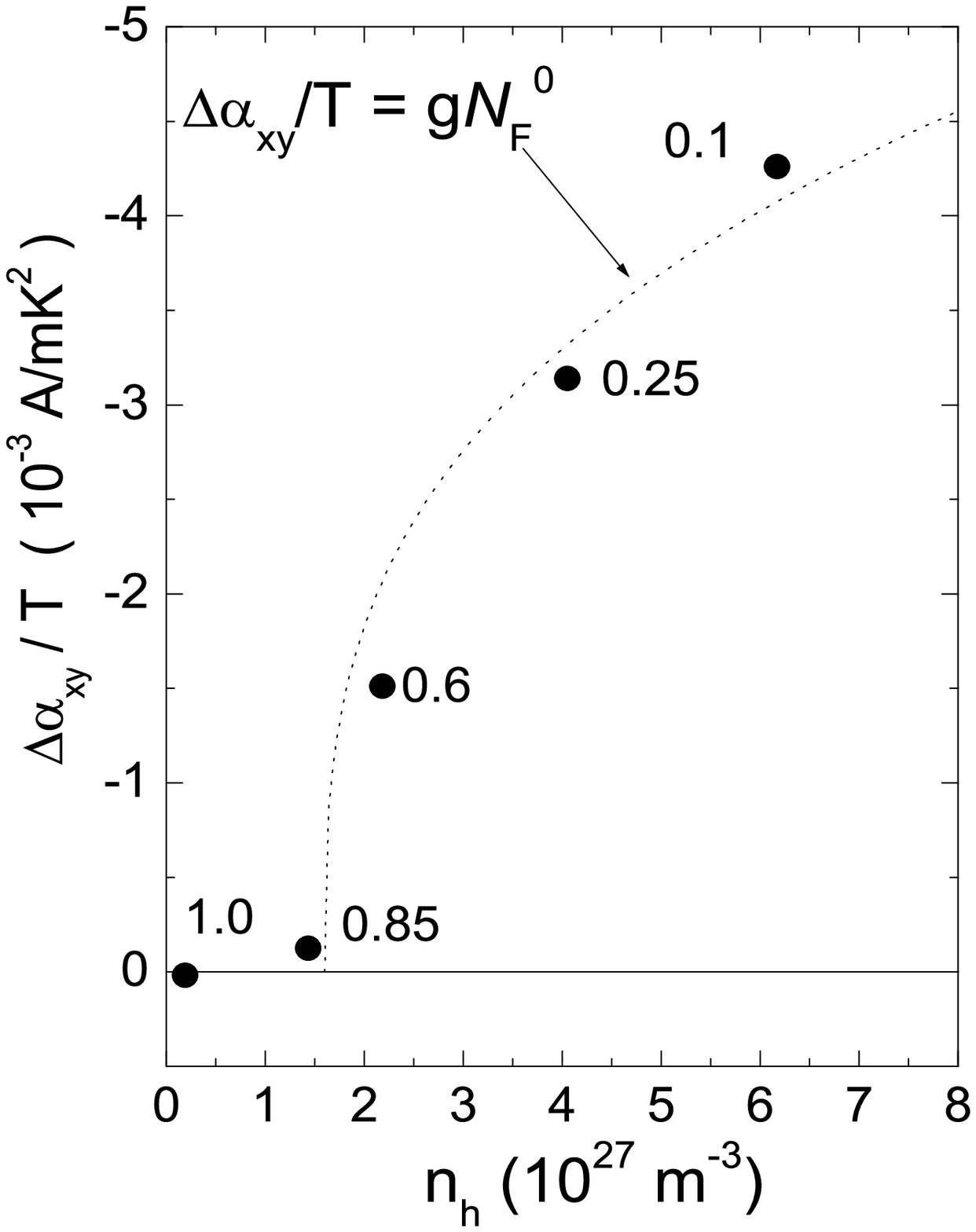}
\caption{\label{axyk}  Plot of $b(x) = \Delta\alpha_{xy}/T$ against $n_h$ showing that above the threshold doping ($x\sim$ 0.85), $b(x)$ increases as a %%@
fractional power of $n_h-n_{h0}$.  The dashed line is $\Delta\alpha_{xy}/T = g{\cal N}_F^0$ with $g = 9.77\times 10^{-50}$ in SI %%@
units. 
}
\efig
Finally, the derived curves of $\alpha_{xy}$ vs. $T$ are shown in Fig. \ref{axyT}.  In contrast to the non-monotonic behavior of $e_N$ vs. $x$, $\alpha_{xy}$ %%@
varies linearly with $T$ as $\alpha_{xy} = b(x)T + c$, where the slope $b(x)$ now decreases monotonically as $x$ increases from 0.1 to 1.0 (Fig. %%@
\ref{axyT}).  In all samples except $x$ = 0.25, the parameter $c$ -- probably extrinsic in nature -- is close to zero within our accuracy.  The dependence of %%@
the parameter $b(x) = [\alpha_{xy}(T)-\alpha_{xy}(0)]/T$ on these two quantities is of main interest.  Figure \ref{axyT}b compares how $b(x)$ and $n_h$ -- %%@
determined~\cite{Lee} from $\rho_{xy}$ above $T_C$ -- vary with $x$.  Whereas, at small $x$, the decrease in $b(x)$ seems to match that of $n_h$, %%@
$b(x)$ falls much faster to zero at large $x$.  

A striking relation between them becomes apparent if we plot one against the other.  Figure \ref{axyk} shows that, above a threshold doping near 0.85, %%@
$b(x)$ increases as a fractional power of $n_h- n_{h0}$ with $n_{h0}$ a threshold density.  This is consistent with $b(x)$ increasing like the density of %%@
states (DOS), viz. $b(x)\sim {\cal N}_F$.  The DOS for the free-electron gas ${\cal N}_F^0$ (dashed curve) has a slightly stronger curvature than the data.  %%@
Interestingly, the occurence of the threshold doping at $x$ = 0.85 accounts well for a puzzling change in the resistivity behavior when $x$ exceeds 0.85 %%@
(Fig. \ref{Stan}c).  In general, slowly increasing $x$ causes the resistivity profiles to change systematically, reflecting slight decreases in both $n_h$ and %%@
$\ell_0$ (mean-free-path).  However, between 0.85 and 1.0, the change is sudden and striking.  At $x$ = 0.85, $\rho$ is $T$ independent below 100 K %%@
consistent with a disordered metal.  By contrast, at 1.0, $\rho$ rises monotonically with decreasing $T$ (Fig. \ref{Stan}c).  Between 300 and 4.2 K, $\rho$ %%@
increases from 6.3 to 32 m$\Omega$cm.  At low $T$, conductivity proceeds by hopping between strongly localized states in an impurity band.  Figure %%@
\ref{axyk} confirms that we reach the extremum of the hole band near $x$ = 0.85.  Further removal of carriers ($x\rightarrow 1$) affects states within the %%@
impurity band.

Knowing $n_h$ and $\rho$ at each $x$, we may determine the mean-free-path $\ell_0$ in the impurity-scattering regime.  Between $x$ = 0.1 and 1.0, %%@
$\ell_0$ decreases by a factor of 40.  This steep decrease has no discernible influence on $b(x)$.  Combining these factors then, we have $\alpha_{xy} = %%@
gT{\cal N}_F$ where $g$ is independent of $\ell_0$.  We may boil down $\alpha_{xy}$ to the measurement of an ``area" ${\cal A}$ by %%@
writing
\be 
\alpha_{xy} = {\cal A}\frac{ek_B^2T}{\hbar}{\cal N}_F, \quad\quad\quad(T\ll T_C).
\label{aa}
\ee
with $k_B$ Boltzmann's constant and $e$ the electron charge.  The value of $g$ gives ${\cal A} = 33.8$ \AA$^2$ if we assume ${\cal N}_F\sim {\cal %%@
N}_F^0$.  As the anomalous Hall heat current produced by $\bf E||\hat{x}$ is $J^Q_y = \alpha_{yx}TE$, it shares the simple form in Eq. \ref{aa}.  The ratio %%@
$J^Q_y/J_y \sim T^2$, as expected for a Fermi gas.

We briefly sketch the anomalous-velocity theory~\cite{Karplus,Niu,Nagaosa}.  In a periodic lattice, the position operator for an electron is the sum $\bf x = %%@
R + X(k)$ where ${\bf R}$ locates a unit cell, while $\bf X(k)$ locates the intracell position~\cite{Adams}.  A finite $\bf X(k)$ implies that $\bf x$ does not %%@
commute with itself.  Instead, we have~\cite{Adams} $[x_j, x_k] = i\epsilon^{jkm}\Omega_m$ with $\epsilon^{jkm}$ the antisymmetric tensor, which implies %%@
the uncertainty relation $\Delta x_j\Delta x_k\sim \Omega$.  The `Berry curvature' $\bf \Omega(k) = \nabla_{k}\times X$ is analogous to a magnetic field in %%@
$\bf k$ space~\cite{Berry}.  In the presence of $\bf E$, $\bf \Omega$ adds a term that is the analog of the Lorentz force to the %%@
velocity $\bf v_k$, viz.  
\be
\hbar{\bf v_k} =\mbox{\boldmath{$\nabla$}}\epsilon({\bf k}) - {\bf E\times \Omega(k)}.
\label{v}
\ee
The anomalous velocity term in Eq. \ref{v} immediately implies the existence of a spontaneous Hall current ${\bf J}_H = -2e\sum_{\bf k} f^0_{\bf k} {\bf %%@
E\times \Omega(k)}$ where $f^0_{\bf k}$ is the unperturbed distribution~\cite{Karplus,Adams,Niu,Nagaosa,Yao}.  The unconventional form of the current -- %%@
notably the absence of any lifetime dependence -- has made the KL theory controversial for decades~\cite{Hurd,Smit}.  However, strong support has been %%@
obtained from measurements of Lee \etal~\cite{Lee} showing that the normalized AHE conductivity $\sigma'_{xy}/n_h$ in $\rm CuCr_2Se_{4-x}Br_x$ is %%@
unchanged despite a 1,000-fold increase in $\rho$.

In general, the off-diagonal term $\alpha_{xy}$ is related to the derivative of $\sigma_{xy}$ at the chemical potential $\mu$, viz.
$\alpha_{xy} = \frac{\pi^2}{3}\frac{k_B^2T}{e} \left[\frac{\partial \sigma_{xy}}{\partial\epsilon}\right]_{\mu}$~\cite{Ong}.
Using the result~\cite{Lee} that $\sigma'_{xy}$ is linear in $n_h$ but independent of $\ell_0$, and $[\partial n_h/\partial\epsilon]_{\mu} = {\cal N}_F$, we %%@
see that $\alpha_{xy}\sim T{\cal N}_F$, consistent with Eq. \ref{aa}.  [By contrast, we note that the skew-scattering model~\cite{Smit} would predict %%@
$\sigma'_{xy}\sim n_h\ell_0$ and $\alpha_{xy}\sim T{\cal N}_F\ell_0$.]   

Finally, ${\cal A}$ in Eq. \ref{aa} has the value 34 $\mathrm{\AA}^2$.  If Eq. \ref{v} is indeed the origin of $\alpha_{xy}$, ${\cal A}$ must be roughly the scale %%@
of $\Omega \sim\Delta x_j\Delta x_k$.  Hence the value ${\cal A} \sim\frac13$ the unit-cell area seems reasonable (the lattice spacing here is 10.33 \AA).  
While a quantitative comparison requires knowledge of $\bf \Omega(k)$ over the Brillouin zone, the simple form of Eq. \ref{aa} seems to provide valuable %%@
insight on the anomalous heat current.

A previous calculation of the Nernst coefficient was based on the ``side-jump" model~\cite{Berger}.  On scattering from an impurity, the carrier suffers a %%@
small sideways displacement $\delta$ to give on average $\tan\theta_H = \delta/\ell_0$.  This was used to derive $Q_s\sim T(k_F\ell)^{-1}$.  In our %%@
experiment, $k_F\ell$ falls monotonically, with increasing $x$, while $e_N$ rises to a broad maximum near 0.25 before falling.  Hence our experiment is in %%@
essential conflict with the side-jump model.  From earlier experiments~\cite{Kondorskii}, an empirical form $Q_s = -(a + b'\rho)T$ has been inferred %%@
($a,b'$ are constants).   This is not borne out in our data.

Combining Nernst, Hall and thermopower experiments on the ferromagnet $\rm CuCr_2Se_{4-x}Br_x$, we have determined how $\alpha_{xy}$ (hence %%@
${\bf J}^Q$) changes as a function of $n_h$ and $\tau$.  At low $T$, we find that $\alpha_{xy}$ follows the strikingly simple form $\alpha_{xy}\sim {\cal %%@
A}T{\cal N}_F$, consistent with the anomalous-velocity theory for the AHE (Fig. \ref{aa}).  In addition, a direct relation (Eq. \ref{beta}) between $M$ and %%@
$\alpha_{xy}$ is observed in the paramagnetic regime above $T_C$.

We acknowledge support from the U.S. National Science Foundation (Grant DMR 0213706).
\newline
{\emph{$^*$Permanent address of S. W. : Center for Crystal Science and Technology, University of Yamanashi, 7 Miyamae, Kofu, Yamanashi 400-8511, %%@
Japan}


\begin{thebibliography}{99}
\bibitem{vortex}  The value of $\beta$ for vortex flow in superconductors is larger than in ferromagnets by $\sim 10^5$ (see Eq. \ref{beta}).  This reflects a %%@
fundamental difference in how magnetization couples to the Nernst signal in the 2 cases.
\end{thebibliography}
\end{document}